\documentclass[twocolumn,aps,prb,unsortedaddress,superscriptaddress]{revtex4}
\usepackage{graphicx}
\usepackage{amssymb}
\usepackage{amsmath}
\usepackage{amscd}
\usepackage{setspace}
\usepackage{graphicx}
\usepackage{epsfig}
\usepackage[sc]{subfigure}
\usepackage{natbib}
\usepackage{helvet}
\usepackage{enumerate}
\usepackage{dcolumn}% Align table columns on decimal point
\usepackage{bm}% bold math
\usepackage{color}%
\usepackage{ulem}%
\usepackage{sidecap}
\usepackage{color}
\usepackage{lineno}

\begin{document}

\title{Real space first-principles derived semiempirical pseudopotentials applied to tunneling magnetoresistance} 

\author{K. H. Bevan$^*$}
\affiliation{NSF Network for Computational Nanotechnology,\\
 Purdue University, West Lafayette, IN, 47907, USA}
\affiliation{Materials Science and Technology Division, Oak Ridge National Laboratory, Oak Ridge, Tennessee 37831, USA}
\affiliation{Centre for the Physics of Materials and Department of Physics, McGill 
University, Montreal, PQ, H3A 2T8, Canada}
\author{Tony Low}
\affiliation{NSF Network for Computational Nanotechnology,\\
 Purdue University, West Lafayette, IN, 47907, USA}
\author{H. Guo}
\affiliation{Centre for the Physics of Materials and Department of Physics, McGill 
University, Montreal, PQ, H3A 2T8, Canada}

\date{\today}

\widetext

\begin{abstract}

In this letter we present a real space density functional theory (DFT) localized basis set semi-empirical pseudopotential (SEP) approach.   The method is applied to iron and magnesium oxide, where bulk SEP and local spin density approximation (LSDA) band structure calculations are shown to agree within approximately 0.1 eV.   Subsequently we investigate the qualitative transferability of bulk derived SEPs to Fe/MgO/Fe tunnel junctions.   We find that the SEP method is particularly well suited to address the tight binding transferability problem because the transferability error at the interface can be characterized not only in orbital space (via the interface local density of states) but also in real space (via the system potential).   To achieve a quantitative parameterization, we introduce the notion of \textit{ghost} semi-empirical pseudopotentials extracted from the first-principles calculated Fe/MgO bonding interface.   Such interface corrections are shown to be particularly necessary for barrier widths in the range of 1 nm, where interface states on opposite sides of the barrier couple effectively and play a important role in the transmission characteristics.   In general the results underscore the need for separate tight binding interface and bulk parameter sets when modeling conduction through thin heterojunctions on the nanoscale.

\end{abstract}

\maketitle

% ====================================================================

\section{Introduction}
\label{sec1}

% Why we are interested in Fe/MgO/Fe system
Recently, first-principles theoretical predictions of crystalline Fe/MgO/Fe tunneling magnetoresistance (TMR) on the order of several hundred percent or more\cite{ButlerTMR,MathonTMR} were confirmed in series of notable experiments.\cite{JapanTMR_MgO,AlmadenTMRMgO}   With optimization efforts continuing, this dramatic TMR enhancement has placed magnetic tunnel junction (MTJ) devices in a unique position to revolutionize memory, magnetic sensor, and computing technologies.\cite{SwitchingSpeedDiscussionAndTMR,IndustryMgORev} 

% Why interface of Fe/MgO/Fe is important for device physics
The large tunneling magnetoresistance of crystalline Fe/MgO/Fe junctions can be understood in terms of the symmetry of the MgO crystal, which allows states with $\Delta_{1}$ symmetry to transmit efficiently through the band gap of MgO  while states of $\Delta_{2/5}$ symmetry decay rapidly.\cite{ButlerTMR}  Near the Fermi energy, the Fe majority and minority states are primarily of $\Delta_{1}$ and $\Delta_{5}$  symmetry respectively.   Therefore the MgO barrier acts as a spin filter, resulting in half-metallic like conduction between $\Delta_{1}$ states on opposite sides of the barrier.  Studies have shown that only a single crystalline Fe layer adjacent to MgO is sufficient to produce most of the TMR observed in thicker Fe/MgO/Fe devices.\cite{GermanyMgORev,heiliger06}  Recently, it was also suggested that spin torque transfer largely occurs at the Fe/MgO interface.\cite{heiliger08}  Therefore it is essential that any Fe/MgO/Fe MTJ device transport model correctly capture the physical properties of the Fe/MgO interface.

% We need to strike a compromise i.e. DFT, empirical TB, pseudopotential based empirical methods. How? Why?
From a computational perspective, the scalability of density functional theory (DFT) in magnetic metals presents serious limitations.\cite{SpinConvergence, SiestaOrderN}  For example, the study of spin torque\cite{SayeefSpinTorque}  and TMR through large scale  MTJ cross sections interspersed with magnetic impurities and/or crystal defects \cite{ExpDefectPaperTMR, TheoryDefectPaperTMR,  AliSpin} would be computationally prohibitive. Scalability is particularly problematic in non-collinear magnetic tunnel junction systems, where the calculation convergence time can be prodigious \cite{SpinConvergence, WaldronNanotechnology} due to the additional spin degree of freedom. Furthermore, the tendency of DFT to underestimate semiconductor and insulator band gaps limits its capability to quantitatively model device transport characteristics. For example in magnetic tunnel junctions, the commonly applied local spin density approximation (LSDA)\cite{ButlerTMR,  WaldronNanotechnology, SelfInteractionCorrectedMTJ, Martin}  significantly underestimates the MgO band gap and therefore over estimates the tunneling current and spin-torque transfer. In light of these concerns, we are motivated to study the applicability of employing the semi-empirical pseudopotential method \cite{ZungerSEP} in the context of Fe/MgO/Fe magnetic tunnel junctions (MTJs). 

% Briefly discuss SEP method
The semi-empirical pseudopotential method (herein known by the abbreviation SEP) assumes that the Hartree and exchange-correlation potential interaction between electrons in a crystal lattice can be well approximated by an angular dependent or spherically symmetric potential situated at each atomic site. In its simplest form, where we assume a spherically symmetric SEP, the approach is analogous to the atomic sphere approximation. \cite{Martin}  The SEP approximation was first applied to plane wave calculations and benchmarked with respect to the bulk properties of Si and CdSe.\cite{ZungerSEP}  The implementation was later scaled up and applied to the study of quantum dot systems \cite{ZungerDot} possessing a large number of atoms. By optimizing the SEP parameter set one is able to correct the band gap of the modeled material while maintaining DFT wavefunction accuracy.\cite{ZungerSEP} The latter feature is of utmost importance in Fe/MgO/Fe tunnel junctions, where wavefunction symmetry plays a pivotal role in the device transport characteristics.  Furthermore, this approximation removes the need for a self-consistent convergence loop and therefore allows for the study of much larger systems. The method is also appealing from a tight-binding perspective, \cite{CerdaEHT, BoykinTB}  since it offers the same computational advantages and yet is able to rapidly produce an accurate parameterization without employing sophisticated optimization algorithms.\cite{KlimeckGenetic}       

Building upon previous theoretical studies, \cite{MacLarenTMR, ButlerTMR, ZhangTMR, MathonTMR} we examine the applicability of employing semi-empirical pseudopotentials  (SEPs) \cite{ZungerSEP} for the study of Fe/MgO/Fe tunnel junctions within a real space localized basis set calculations (rather than plane wave calculations \cite{ZungerSEP}). The discussion is divided in two parts.  Firstly, the SEP extraction method is described in detail.  Secondly, we evaluate the SEP method with respect to bulk, interface and thin barrier parameterizations. 

The method is first benchmarked against bulk Fe and bulk MgO LSDA band structure calculations.  Subsequently, we show that the bulk derived SEPs are \textit{unable} to quantitatively capture the LSDA derived Fe/MgO/Fe interface and thin barrier tunneling characteristics. \cite{ButlerTMR, WaldronNanotechnology} To overcome this shortcoming we therefore introduce a separate interface parameterization through the concept of \textit{ghost} semi-empirical potentials localized between the Fe and MgO interface atoms. With these interface corrections, we are then able to quantitatively capture DFT tunneling through thin barriers. It is shown that an accurate interface parameterization is required for barrier widths in the range of 1 nm, where  interface states on opposite sides of the barrier can couple strongly.    We also evaluate the transferability and importance of MgO barrier band gap corrections with respect to the total barrier transmission.   In general the results underscore the need for separate interface and bulk parameterization sets when modeling electron transport through thin tunnel junctions.

% ====================================================================

\section{Method}
\label{sec2}

We briefly outline our simulation method in this section in two parts.   Firstly, we outline the chosen local atomic orbital DFT method.  Secondly, we discuss the real space semi-empirical pseudopotential approximation applied in this work.   The self-consistent non-equilibrium green's function (NEGF) DFT transport method applied in this work has been discussed extensively in previous publications.\cite{WaldronPRL, WaldronNanotechnology}

 \subsection{Local Atomic Orbital DFT Method}
\label{sec2.1}

The local atomic orbital pseudopotential DFT time independent Hamiltonian can be expressed as,
\begin{align}
\hat{H} = -\frac{1}{2}\nabla^2 + \hat{V}_{ps}({\bf r})+V^{H}({\bf r})+ V^{XC}[\rho({\bf r})],  
\end{align}
where $\hat{V}_{ps}$ is the pseudopotential term, $V^{H}$ is the Hartree term, $V^{XC}$ is the exchange-correlation potential term and $\rho$ is the system charge density.  
We may expand the pseudopotential expression further into local and non-local terms following the Klienman-Bylander prescription,\cite{PseudopotentialKB}
\begin{align}
\hat{V}_{ps}({\bf r}) &= \hat{V}^{nloc}_{ps}({\bf r}) +  V^{loc}_{ps}({\bf r}) \\
&=  \hat{V}^{nloc}_{ps}({\bf r})
+ \sum_{\alpha = 1}^{N} v_{ps,\alpha}(|{\bf r} - {\bf r}_{\alpha}|).
\end{align}
 where $\alpha$ is the atomic index and ${\bf r}_{\alpha}$ is a summation taken across the pseudopotentials of each atomic position.   However, $V^{loc}_{ps}({\bf r})$ is usually long ranged (which reduces the sparsity of the Hamiltonian) and therefore also computationally problematic.   Thus, we screen $V^{loc}_{ps}({\bf r})$ \cite{SIESTA} by populating the orbitals of the isolated atom and arrive at a short ranged neutral atom potential, $V^{NA}(r)$, for each atomic species.   The preferred local atomic orbital Hamiltonian is then written as
\begin{align}
\hat{H} = &-\frac{1}{2}\nabla^2 + \hat{V}_{ps}^{nloc}({\bf r}) +  \sum_{\alpha = 1}^{N} V^{NA}_{\alpha}(|{\bf r} - {\bf r}_{\alpha}|)& \nonumber \\
&+ \delta V^{H}({\bf r})+ V^{XC}[\rho({\bf r})],&
\end{align}
such that the modified Hartree term is given by $\nabla^2  \delta V^H({\bf r}) = -4\pi \delta \rho({\bf r})$.   We define $\delta \rho({\bf r}) = \rho({\bf r}) - \sum_{\alpha}  \rho^{atom}_{\alpha}(r)$ where $\rho^{atom}_{\alpha}$ is the neutral atom charge arrived at by populating the orbitals of an atomic species.

\subsection{Semi-empirical Pseudopotentials}
\label{sec2.1}
\subsubsection{Extracting SEPs from real space DFT calculations}

In its most basic form, the semi-empirical pseudopotential approximation \cite{ZungerSEP} assumes that all local terms may approximated by a spherically symmetric local potential around each atom.  This is objective is partially accomplished by including  $V^{NA}(r)$ but to arrive at a proper spherical potential at each atomic site we must also reduce $\delta V_{H}$ and  $V_{XC}$ such that,
\begin{align}
 \delta V^{H}({\bf r}) + V^{XC}[\rho({\bf r})]
\approx \sum_{\alpha = 1}^{N} v_{\alpha}(|{\bf r} - {\bf r}_{\alpha}|).
\label{SEP_Eq}
\end{align}
where $v_{\alpha}$ is the spherical approximation to the self-consistent Hartree and exchange-correlation terms for atom $\alpha$.   The full semi-empirical pseudopotential is given by $V_{\alpha}^{SEP}(r) = v_{\alpha}(r) + V_{\alpha}^{NA}(r)$. 
The approach is similar in spirit to the atomic sphere approximation applied in the muffin-tin orbital method.\cite{Martin}  Although not done here, angular dependence may be introduced to the SEP term.
This leads to a revised Hamiltonian operator,
\begin{align}
\hat{H}_{SEP} =
-\frac{1}{2}\nabla^2
+ \hat{V}^{nloc}_{ps}({\bf r})
+  \sum_{\alpha = 1}^{N} V^{SEP}_{\alpha}(|{\bf r} - {\bf r}_{\alpha}|)
\end{align}
which does not require a self-consistent loop to solve since there is no interdependence
between the semi-empirical pseudopotentials and the charge density.   The term ``semi-empirical" is applied because these potentials are initially derived from first-principles calculations and then fitted to experimental data if required -- in this work to overcome the LSDA band gap underestimation.     

Finally, we would like to extract the SEP for each atomic species from a self-consistent DFT
calculation.   Let us assume that the spherical approximation to the self-consistent Hartree and exchange-correlation terms, $v_{\alpha}(r)$, goes to zero beyond a cutoff radius of $r_c$ -- which is not necessarily equivalent to the cutoff radius of $V_{\alpha}^{NA}(r)$.    Note that the zero potential condition outside the cutoff radius may be adjusted, for example by adding a positive offset to the real space DFT potential.   Within the cutoff radius we may define a complete orthonormal basis \cite{ArfkenWeber} to represent $v_{\alpha}(r)$ such that
\begin{align}
\phi_n(r)
=  
\left\{
\begin{array}{lr}
\frac{1}{\sqrt{2\pi r_c}} \frac{sin(n\pi r/ r_c)}{r}&   r  \le r_c  \\
0   &   r > r_c  \\
\end{array}
\right.
\end{align}
and
\begin{align}
 v_{\alpha}(r)=\sum_{n=1}^M c^{\alpha}_n \phi_n,
 \label{SEP_Sum_Eq}
 \end{align}
where the potential is represented by a linear expansion of the zeroth order spherical Bessel function -- the eigenfunctions of an electron with no angular momentum trapped in an infinite spherical well of radius $r_c$.  Note that higher order spherical Bessel functions are not able to capture a non-zero system potential at the atomic origin.
To solve for the coefficients $c^{\alpha}_n$ we substitute Eq. (\ref{SEP_Sum_Eq}) into Eq. (\ref{SEP_Eq}) and construct a linear equation, with $N \times M$ unknowns, by integrating both sides through with $\phi_m$ centered at atomic species $\beta$,
\begin{align}
&\int  \phi_m(|{\bf r} - {\bf r}_{\beta}|) ( \delta V^{H}({\bf r}) + V^{XC}[\rho({\bf r})] )
 d{\bf r}& \nonumber \\
&=  \sum_{\alpha = 1}^{N} \sum_{n=1}^N c^{\alpha}_n
\int  \phi_m(|{\bf r} - {\bf r}_{\beta}|)
  \phi_n(|{\bf r} - {\bf r}_{\alpha}|)d{\bf r}.&
\label{SEP_Integral_Eq}
\end{align}
In Eq. (\ref{SEP_Integral_Eq}) we have forced an equality between the self-consistent DFT local potential and the spherical SEP approximation to that potential.  By further considering all Bessel functions in our SEP expansion we obtain linear system of equations, ${\bf V} = [S]{\bf c}$, which may be written as
\begin{align}
 \left[ \begin{matrix} V_{1}^{1} \\
                              V_{2}^{1} \\
                              \vdots \\
                              V_{M}^{N}
                               \end{matrix} \right]
=
 \left[ \begin{matrix} S_{11}^{11} & S_{12}^{11} & ... & S_{1M}^{1N}  \\
                                S_{21}^{11} & S_{22}^{11} &  \\
                                \vdots &   & \ddots  &  \\
                                S_{M1}^{N 1} &  &  & S_{MM}^{N N}
                                \end{matrix} \right]
 \left[ \begin{matrix} c_{1}^{1} \\
                              c_{2}^{1} \\
                              \vdots \\
                              c_{M}^{N}
                               \end{matrix} \right]
\label{SEP_Matrix_Eq}                           
\end{align}
where $V_{m}^{\beta}$ denotes a Bessel integral over  $\delta V^{H}({\bf r}) + V^{XC}[\rho({\bf r})]$ and $S_{mn}^{\beta\alpha}$ denotes an overlap integral on the right had side of Eq.~(\ref{SEP_Integral_Eq}).   The coefficients $c_{n}^{\alpha}$  are then directly arrived at by matrix inversion.

\subsubsection{Extension to bulk systems}

In bulk periodic systems, we need to consider the periodicity of the system potential when solving for the SEP Bessel coefficients.   If we take the left hand side of Eq.~(\ref{SEP_Integral_Eq}) and integrate through the SEP Bessel functions of a given unit cell, the periodic potential  ($\delta V^{H}({\bf r}) + V^{XC}[\rho({\bf r})]$) over which the integral is performed will have contributions not only due to the SEPs of the unit cell which we have selected but also due to the SEPs of neighboring unit cells.    We can address this issue by adopting a supercell tight binding description of bulk periodicity, where beyond twice the maximum SEP cutoff radius the interaction between a unit cell and its bulk neighbors is assumed to go to zero.    In this manner, the SEP Bessel integrals on the left hand side of  Eq.~(\ref{SEP_Matrix_Eq}) are performed only for the central unit cell in our supercell.    However, the SEP coefficients must be the same for all unit cells.   Therefore, the SEP matrix overlap matrix on the right  hand side of Eq. (\ref{SEP_Matrix_Eq}) is expanded into a summation of the SEP overlap matrices between the central unit cell and all neighboring unit cells within the supercell.   The revised unit cell SEP integral equation is then written as, 
\begin{align}
&\int  \phi_m(|{\bf r} - {\bf r}_{\beta}|) ( \delta V^{H}({\bf r}) + V^{XC}[\rho({\bf r})] )
 d{\bf r}& \nonumber \\
&=   \sum_{\alpha = 1}^{N} \sum_{n=1}^N c^{\alpha}_n
\sum_{{\bf R}}
\int  \phi_m(|{\bf r} - {\bf r}_{\beta}|)
  \phi_n(|{\bf r} - {\bf r}_{\alpha}-{\bf R}|)d{\bf r}.&
\label{Revised_SEP_Integral_Eq}
\end{align}
such that ${\bf R}=n_1{\bf R}_1+n_2{\bf R}_2+n_3{\bf R}_3$, where ${\bf R}_{1,2,3}$ are the translation vectors of the unit cell and $n_{1,2,3}$ are integers.

\subsubsection{Extension to collinear spin polarized systems}

Thus far we have only outlined the SEP extraction procedure for spin independent calculations. When modeling collinear spin polarized systems, separate SEPs are extracted  for the majority and minority spin electrons.  For majority spin up electrons we simply set $\delta V^{H} + V^{XC \uparrow}[\rho({\bf r})]
\approx \sum_{\alpha = 1}^{N} v_{\alpha}^{\uparrow}(|{\bf r} - {\bf r}_{\alpha}|)$ in Eq. (\ref{SEP_Integral_Eq}), and  solve for the $c_{n}^{\alpha\uparrow}$ zeroth order Bessel coefficients following the above prescription.  In the same manner one is able to extract the minority spin down SEP coefficients.

\subsubsection{Band structure fitting methodology}

To fit the SEP band structure of a given bulk material, the Jacobian matrix $[J]$ of target band structure points $p_j$ is computed with respect to the material SEP coefficients $c_i$ such that $[J]_{ij}=\partial p_j/  \partial c_i$.   For example, to correct the band gap of insulators or semiconductors, the valance and conduction band energies at symmetry $k$-points can be taken as targets to be raised or lowered.     After computing the Jacobian, a new set of coefficients is computed via
\begin{align}
{\bf c}^{new} = {\bf c} + [J]^{-1}{\bf p}
\label{SEP_optimization_Eq}                           
\end{align}
where $[J]^{-1}$ is the pseudo-inverse of the Jacobian if the matrix is not square and ${\bf p}$ is the vector between the existing band structure target values (derived from ${\bf c}$) and the desired band structure target values.   The process is iterated by setting ${\bf c}={\bf c}^{new}$ and recalculating the Jacobian, until the vector ${\bf p}$ approaches a small tolerance value (say 0.1 eV per target value).

% ====================================================================

\section{SEP Electronic Structure for Iron and Magnesium Oxide}
\label{sec3}

% -------------------------------------------------------------------------------------------------------------

\subsection{Bulk Fe and MgO}
\label{sec3.1}

To insure quantitative transport calculations,  in agreement with existing DFT methods, the semi-empirical pseudopotential approximation must be benchmarked against self-consistent results.  Therefore, we begin by examining the accuracy of the semi-empirical pseudopotential method detailed in Sec.~\ref{sec2} as applied to bulk iron and bulk magnesium oxide LSDA calculations.  

\begin{figure}[t]
\centering
\includegraphics[width=3in,height=4in]{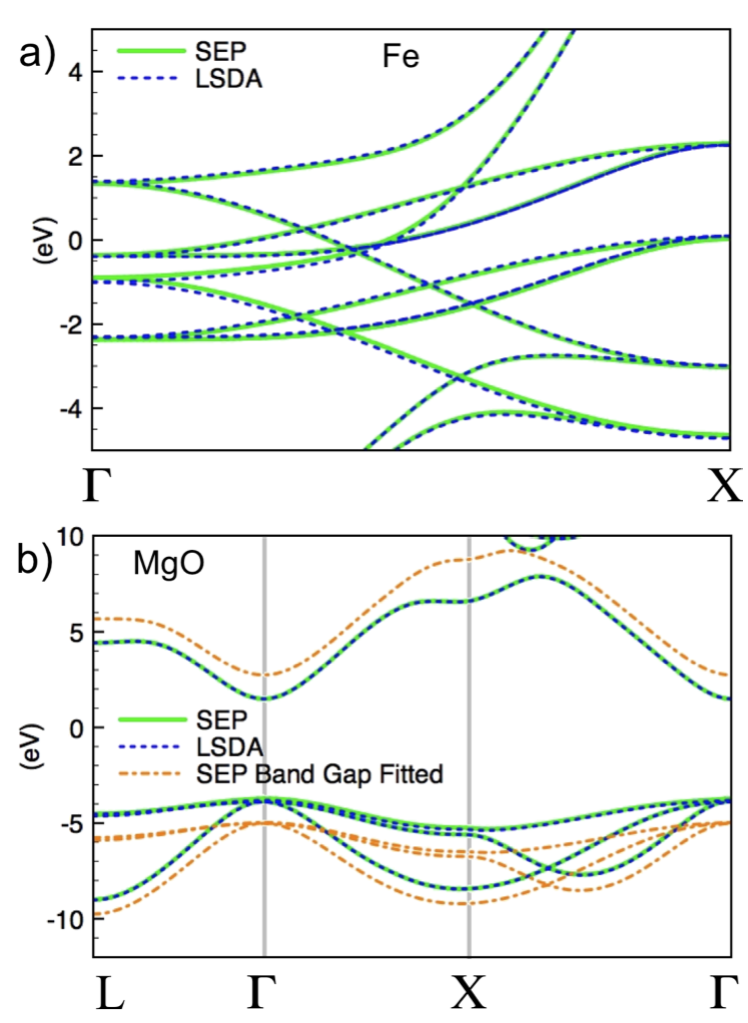}
\caption{Bulk band structure of Fe and MgO.  Subfigure a) provides the Fe (100) crystal bulk band structure with the Fermi energy is situated at $0$ eV.  Subfigure b) provides the strained MgO bulk band structure with the Fermi energy positioned to match that of MgO sandwiched between two Fe(100) slabs.   The LSDA calculated band structure is shown as a dashed blue line.  The LSDA SEP fit is shown in as a solid green line.  The modified LSDA SEP result fitted to the bulk MgO band gap of 7.7 eV\cite{MgOBandGap} is shown as a dot-dashed gold line.}
\label{fig1}
\end{figure}

The band structure of bulk Fe in the (001) direction is presented in Fig.~\ref{fig1}a, where (001) is the direction of electron transport through Fe/MgO/Fe tunneling barriers. \cite{WaldronPRL, ButlerTMR, JapanTMR_MgO, AlmadenTMRMgO}  The lattice constant of Fe is set at 2.87 \AA\ \cite{AshcroftMermin} and a long range double-$\zeta$ polarized basis set \cite{WaldronPRL} is employed in all calculations.   The SEP cut off radius is set to 5 Bohr.  The LSDA calculated band structure (presented as a dashed blue line) in Fig.~\ref{fig1}a can be seen to agree quite well with the SEP calculated band structure (presented as a green line).    The mean margin of error between the two band structure calculations is approximately 0.1 eV.   

The band structure of strained bulk MgO is presented Fig.~\ref{fig1}b.  The MgO lattice constant is set at 4.21 \AA\ in the (001) transport direction.\cite{InterfaceTheoryTMR1,MgOLatticeConstant}   However the (100) and (010) directions are strained by 3.8 \%, to 4.05 \AA, in order to lattice match bulk Fe (see the two probe Fe/MgO/Fe calculations in Sec.~\ref{sec3.2}).  A double-$\zeta$ polarized basis set \cite{WaldronPRL} is employed in all calculations.  The Mg atoms are assigned a basis set a cut off radius of 8 Bohr and the O atoms a cutoff radius of 4.5 Bohr.  The LSDA calculated band structure (presented as a dashed blue line) in Fig.~\ref{fig1}b can be seen to agree quite well with the SEP calculated band structure (presented as a solid green line).    The solid green band structure in Fig.~\ref{fig1}b imposes SEP cutoff radii of 5 Bohr and 4.5 Bohr to Mg and O respectively.  The margin of error between the LSDA and SEP band structures calculations is approximately 0.1 eV.   We have found the same level of SEP fit accuracy can be achieved with the unstrained MgO lattice.   

It is important to note that SEPs with longer cutoff radii (beyond 5 Bohr as demonstrated here) are tenable but often end up sampling not only the potential of the local atom which they are situated on but also the potential of neighboring atoms.   Such SEPs are therefore not even qualitatively transferable to material heterojunctions (for example Fe/MgO/Fe as studied in this work). 
 
To achieve the above fit accuracy, with both bulk Fe and MgO, we applied a LSDA real space grid resolution of 4 points per Bohr (64 points per Bohr$^3$).   To fit the Fe LSDA band structure, 20 Bessel functions for both the up-spin and down-spin SEPs were required, although with shorter range Fe basis sets we have found that as few as 10 Bessel functions are suitable.  To fit the MgO band structure 10 Bessel functions per SEP were required.   Reducing real space the grid resolution reduces the accuracy of the integrals in Eq. (\ref{SEP_Integral_Eq}) and can result in a poor matching between the LSDA and SEP calculated band structures.  Likewise, an insufficient number of Bessel functions in Eq. (\ref{SEP_Sum_Eq}) will result in a poorly constructed SEP.    There is a fine balance between the grid resolution and the number of Bessel functions, as too much of either can raise both the calculation computation time and memory consumption.   

\begin{figure}[t]
\centering
\includegraphics[width=3.25in,height=3.25in]{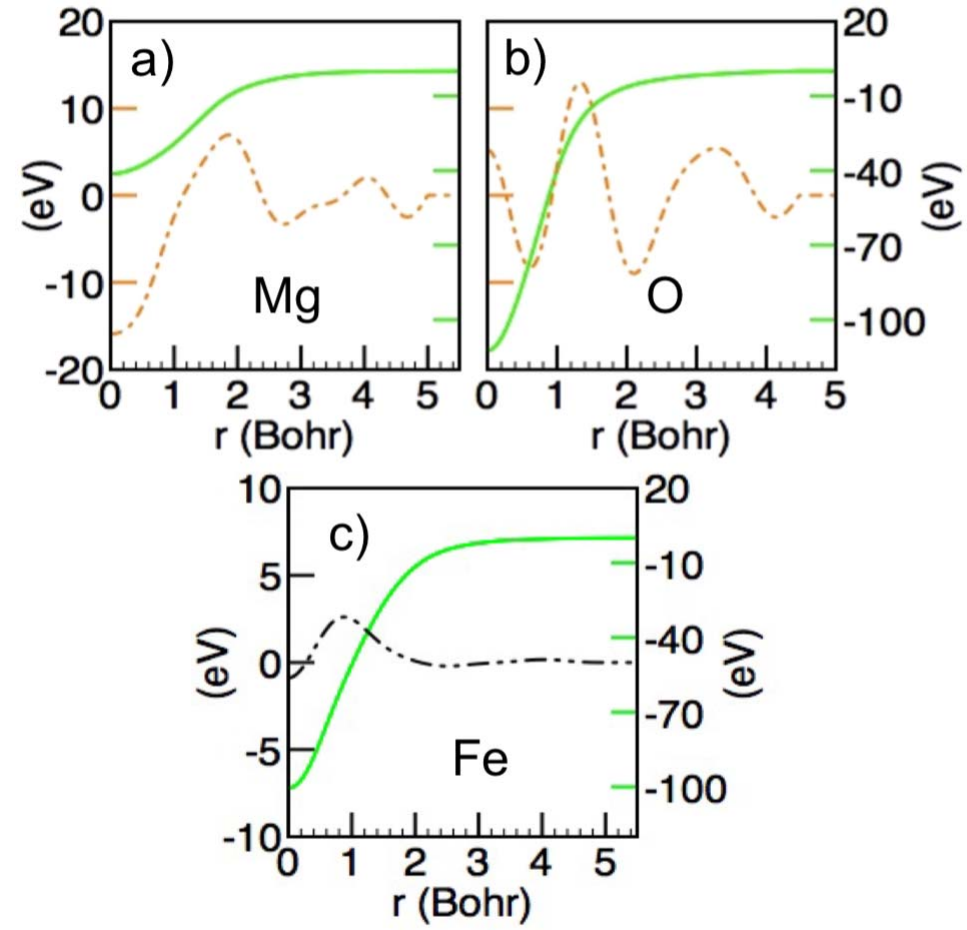}
\caption{Spherical SEPs for each of the elements.  Subfigures a) and b) show the LSDA Mg and O SEPs for MgO in solid green and the MgO band gap fit correction to the LSDA SEPs in dot-dashed gold.   The MgO band gap fit corrections should be read off the left axis and the LSDA SEPs should be read off the right axis.  Subfigure c) shows the LSDA Fe up spin SEP in solid green (read off the right axis) and difference between the Fe down spin and Fe up spin LSDA SEPs ($V^{SEP}_{Fe\downarrow}-V^{SEP}_{Fe\uparrow}$) as a double-dot-dashed black line (read off the left axis).   The vacuum level is set at 0 eV.}
\label{fig2}
\end{figure}

Radial real space plots of the strained bulk MgO and bulk Fe SEPs are presented in Fig.~\ref{fig2}.   The O SEP is much sharper than the Mg SEP (see solid green lines in Figs.~\ref{fig2}a and \ref{fig2}b read off the right axis), and both possess considerable corrections when the MgO band gap is expanded (see dot-dashed gold lines in Figs.~\ref{fig2}a and \ref{fig2}b read of the left axis).    The SEP corrected MgO band structure, fitted to the bulk MgO band gap of 7.7 eV,\cite{MgOBandGap} is plotted as dot dashed gold line in Fig.~\ref{fig1}b.   
We have investigated shorter ranged band gap corrections but have found that they are not able to open the band gap without significantly distorting the band structure.  The Fe SEPs are displayed in Fig.~\ref{fig2}c.  The up spin Fe SEP (see solid green line in Fig.~\ref{fig2}c read off the right axis) and the down spin Fe SEP (see double-dot-dashed black line in Fig.~\ref{fig2}c read off the left axis) differ primarily only with respect to on-site exchange corrections localized at the Fe atomic core.

% -------------------------------------------------------------------------------------------------------------

\subsection{Bulk SEP Transferability to Fe/MgO/Fe Tunnel Junctions}
\label{sec3.2}

% Intro
Given the accurate bulk band structure results presented in the previous section, we now proceed to examine the transferability of bulk derived semi-empirical pseudopotentials to magnetic tunnel junctions.  The bulk SEP zero bias two probe Fe/MgO/Fe TMR ratio, projected density of states (PDOS), and transmission characteristics are shown to qualitatively match first-principles self-consistent NEGF-LSDA results.\cite{WaldronPRL, WaldronNanotechnology}   To obtain a quantitative tunneling barrier parameterization we explicitly identify the real space Hartree and exchange-correlation potential error of bulk SEPs at the Fe/MgO interface, and introduce the notion of \textit{ghost} SEPs to fit and thereby remove the transferability error.  In this regard the SEP tight binding approach is shown to be advantageous as it allows both orbital space and real space characterization of heterojunction interface errors introduced by bulk parameterizations. Non-pseudopotential based tight binding methods,\cite{KlimeckGenetic, CerdaEHT, KienleSi} where the atomic orbitals overlap integrals are used as fitting parameters, do not allow such a systematic characterization of interface transferability errors.

% .........................................................................................................................................

\subsubsection{Fe/MgO/Fe MTJ Geometry}
\label{sec3.2.1}

% Discuss geometry
The Fe/MgO/Fe tunnel junction under investigation consists of 5 MgO layers.\cite{WaldronPRL}  The full NEGF-LSDA device region is shown in Fig.~\ref{fig3}, where the semi-infinite leads are accounted for by self-energy terms in the device Green's function.\cite{WaldronNanotechnology}   We have set the Fe lattice constant in both leads to 2.87 \AA\ -- the MgO transverse lattice constant is also set at 4.05 \AA.   However, the MgO layers are separated by  2.1 \AA\ in the transport direction, matching the bulk MgO lattice constant of 4.2 \AA.\cite{InterfaceTheoryTMR1,MgOLatticeConstant}  The Fe-O bonding distance at the Fe/MgO interface is set at 2.169 \AA.\cite{ButlerTMR}   The unit cell geometry shown in Fig.~\ref{fig3} is periodically repeated infinitely in the transverse $(x,y)$-plane (which lies perpendicular to the tunneling transport $z$-direction).   We have chosen to examine the five layer MgO device geometry, rather than wider or thinner barriers, because at this thickness the bulk MgO band gap reappears in the middle of the barrier.  This allows a proper evaluation of both the interface and bulk properties of the MgO tunneling barrier as approximated by the SEP method.

% .........................................................................................................................................

\subsubsection{Fe/MgO/Fe MTJ Potential Profile Study}
\label{sec3.2.2}

Given that our SEP approach relies upon the fundamental assumption that the potential of a system can be approximated by a summation of local potentials, we begin by comparing the system potential results of two probe SEP and self-consistent NEGF-LSDA calculations.   This  SEP tight binding method allows not only orbital space evaluation, in the form of projected density of states (PDOS) plots, but more importantly real space evaluation of the tight binding Hamiltonian. 

\begin{figure}[t]
\centering
\includegraphics[width=3.25in,height=0.75in]{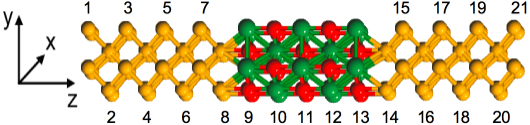}
\caption{Fe/MgO/Fe 5 layer device geometry.  The atomic color index is as follows: iron atoms are colored gold, the magnesium atoms are colored green, and oxygen atoms are colored red. The system is mirror symmetric along the z-axis about layer 11 (the middle of the barrier).}
\label{fig3}
\end{figure}

\begin{figure}[t]
\centering
\includegraphics[width=3.25in,height=2.75in]{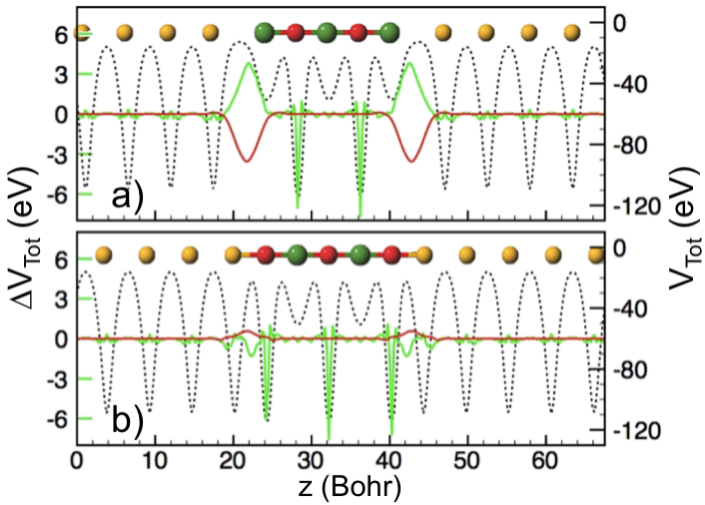}
\caption{The total NEGF-LSDA spin down potential  ($V_{Tot}=V_{Tot}^{LSDA}$) is shown as a dotted black line, with its axis given on the right hand side of the figure (the vacuum level is set at 0 eV).   The bulk LSDA SEP interface error is shown in green ($\Delta V_{Tot} = V_{Tot}^{SEP}-V_{Tot}^{LSDA}$) and the interface spherical SEP correction to the error is shown in red (both potentials are read off the left axis).  Subfigure a) displays the system potential as a linear cut in the $z$-direction through the Mg atom at the Fe/MgO interface.  Subfigure b) displays the system potential as a linear cut in the $z$-direction through the FeO bond at the Fe/MgO interface.  An atomistic cartoon is shown to scale above each potential plot, where a dip in the total potential corresponds to an atomic nuclear position -- Fe atoms are gold, O atoms are red, and Mg atoms are green.  Up spin results are nearly identical.  The SEP MgO band gap corrections (see Figs.~\ref{fig1} and \ref{fig2}) are not included in this comparison.}
\label{fig4}
\end{figure}

In Figs.~\ref{fig4}a and ~\ref{fig4}b potential cuts, through the Mg and O interface atoms respectively, of the MgO two probe geometry (see Fig.~\ref{fig3}) are plotted in the electron transport $z$-direction.    The total down spin potential of our 5 layer Fe/MgO/Fe device is given as a dotted black line (to be read off the right axis) and the corresponding two probe bulk SEP transferability error is plotted in solid green (to be read off the left axis).  The up spin and down spin bulk SEP transferability errors are very similar, therefore in the interest of a concise discussion we include only the down spin results.   Lastly, the  MgO SEP potentials discussed in this section do not include a band gap correction (see Figs.~\ref{fig1} and \ref{fig2}), this issue left to Sec.~\ref{sec3.2.4}.

Away from the interface the SEP potential error, shown in green in Fig.~\ref{fig4} and read off the left axis, is largely flat apart from small oscillations on the Fe and Mg atoms and peaks localized on the O atoms.   The small oscillations away from the interface, can be attributed to the spherical approximation where we have neglected angular variations in the crystal potential about an atom.   The sharp errors localized on each oxygen atom are due the small number of Bessel functions (ten per atom) employed in the bulk MgO fit, which are not able to completely capture the rapid drop in the system potential at the oxygen atomic core.   However, due to their sharp nature these peaks contribute negligibly to the integrated Hamiltonian oxygen onsite energies and therefore can be ignored (see Sec.~\ref{sec3.2.3} for further details).    Immediately away from the interface, the bulk and two probe system potentials agree remarkably well.  However, at the interface the SEP potential error is substantial.   It is important to note that bulk MgO SEP and LSDA two-probe MgO Fermi energies have been aligned via a constant bulk potential shift (see Fig.~\ref{fig1}b).  

The interface bulk SEP error is localized at the FeMg junction, see the solid green line Fig.~\ref{fig4}a, and reaches a maximum of approximately 4 eV (read off the left axis of Fig.~\ref{fig4}a).    Yet, the integrated local atomic orbital matrix Hamiltonian errors\cite{MathonFeO_TMR, MathonTMR} occur on both the Mg interface atoms and the oxygen bonded Fe interface atoms.  On the other hand, the FeO potential cut (in Fig.~\ref{fig4}b) displays relatively little error -- although this error is slightly larger for the up spin system potential.    By including the bulk Fe and bulk MgO Fermi level energy offset in our bulk MgO SEP fit, we have largely compensated for the Hartree potential created by charge redistribution at the FeO interface\cite{ButlerTMR}  (a classical analogue to this would be the built in potential profile of a semiconductor p-n junction).  This offset minimizes the FeO bonding potential error, which can be largely attributed to neglected changes in the exchange-correlation potential.    However, at the FeMg interface the bonding environment changes even more drastically, each Mg interface atom loses one nearest neighbor, and the charge redistribution cannot be approximated by the Hartree potential required to align the heterojunction Fermi energies.   By removing a nearest neighbor at the FeMg interface we violate the spherical symmetry that our SEP fit assumes for the chemical environment and therefore a fundamentally \textit{asymmetric} solution to the SEP approximation is required to overcome the transferability error.

To overcome the bulk transferability error we introduce the concept of \textit{ghost} SEPs, that is SEPs which are not localized at an atomic core but instead situated within the bonding region of the heterojunction interface.    Such ghost SEPs are fitted to cancel the SEP interface transferability error, that is the potential difference between the LSDA two probe calculation and the bulk SEP two probe calculation.    In this manner ghost SEPs allow separate DFT  bulk and heterojunction interface parameterizations, which can be applied independently (for example) to study the device transport properties of various barrier widths, spin torque,\cite{SayeefSpinTorque} and the role of electron and spin defect/impurity scattering within the barrier.\cite{AliSpin}        The interface bulk SEP transferability error is analyzed in further detail in Sec.~\ref{sec3.2.3} and Sec.~\ref{sec3.2.4}, with respect to the two probe interface PDOS and transmission properties.   

The down spin ghost SEP parameterization applied throughout this work is shown as a solid red line (read off the left axis) in Figs.~\ref{fig4}a and \ref{fig4}b.   We have employed two spherically symmetric ghost SEPs per spin (four total) localized along the FeO and FeMg line cuts as shown in Fig.~\ref{fig4}, although higher order angular momentum SEPs may also be applied to capture asymmetry at the interface.   The SEPs plotted in Fig.~\ref{fig4} posses a cutoff radius of 5 Bohr and are composed of 10 Bessel functions.    The Bessel coefficients of the ghost SEPs are arrived at by replacing the terms $\delta V^{H}({\bf r})+ V^{XC}({\bf r})$ in Eq.~\ref{SEP_Integral_Eq} with the two probe potential difference $V_{Tot}^{LSDA}({\bf r})-V_{Tot}^{SEP}({\bf r})$.

% .........................................................................................................................................

\subsubsection{Fe/MgO/Fe MTJ PDOS Transferability Study}
\label{sec3.2.3}

To conceptualize the importance of a proper interface parameterization, let us begin by comparing the PDOS results of two probe SEP and self-consistent NEGF-LSDA calculations.  

\begin{figure}[t]
\centering
\includegraphics[width=3.25in,height=5in]{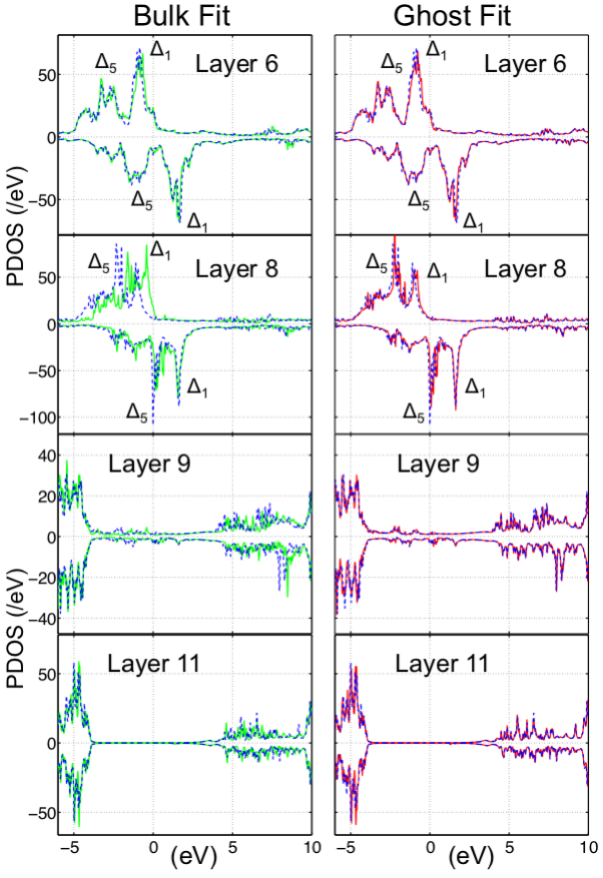}
\caption{Fe/MgO/Fe SEP parallel orientation projected density of states (PDOS) before approaching the interface (layer 5), at the interface (layers 8 and 9), and in the middle of the barrier (layer 11) -- see Fig.~\ref{fig3} for details on the layer numbering.   The Fermi energy is located at 0 eV, the imaginary green's function broadening is set at 25 meV, and the $k$-point sampling set at 8x8 in the transverse Brillouin zone.  The up spin PDOS is shown on the positive axis and the down spin PDOS is shown on the negative axis.  The reference two probe NEGF-LSDA PDOS is shown as a dashed dark blue line in each figure.    The bulk SEP fit (displayed in solid green) applied to the two probe geometry is shown in the first column.  When ghost SEPs (displayed in solid red) are introduced at the interface to correct the bulk Fe and MgO SEP transferability errors, an accurate fit is obtained as shown in the second column.}
\label{fig5}
\end{figure}

The parallel orientation PDOS results at layers 6, 8, 9 and 11 in the Fe/MgO/Fe device are displayed in Fig.~\ref{fig5} -- see the geometry diagram in Fig.~\ref{fig3} for details on the layer numbering.  The two probe NEGF-LSDA PDOS is shown in dashed blue, the bulk SEP PDOS in green, and the ghost SEP PDOS in solid red.  It is important to note the mirror symmetry of the 5 layer Fe/MgO/Fe system, where under zero bias conditions the PDOS at layers 6, 8 and 9 is equivalent to the PDOS at layers 16, 14 and 13 respectively.    

As we transition from deep within the Fe leads towards the MgO tunnel junction, the bulk SEP, NEGF-LSDA, and ghost SEP PDOS calculations agree well.    This agreement holds up until the third Fe layer as measured from the Fe/MgO interface -- see the layer 6 PDOS results in Fig.~\ref{fig5}.   The disagreement reaches a maximum directly at the Fe/MgO interface (see the bulk SEP and ghost SEP fits respectively at layers 8 and 9 in Fig.~\ref{fig5}) where the bulk SEP PDOS begins to diverge from the LSDA PDOS.  Further within the MgO barrier (layer 11 in Fig.~\ref{fig5}) the bulk SEP and ghost SEP parameterizations both capture the LSDA calculated MgO band gap as it begins to reappear.   The agreement between the SEP and LSDA results at layer 11 Fig.~\ref{fig5}, clearly shows that that sharp SEP fit errors located on the O atoms in Fig.~\ref{fig4} do not influence the barrier electronic structure (see the discussion in Sec.~\ref{sec3.2.2}).

Returning to the interface, we see that the layer 8 NEGF-LSDA result  (in dashed dark blue in Fig.~\ref{fig5}) displays the characteristic Fe/MgO interface PDOS  including the minority PDOS Fermi energy resonant peak.\cite{ButlerTMR,InterfaceTheoryTMR2} Yet if we turn our attention to the bulk SEP PDOS layer 8 result (shown in green in Fig.~\ref{fig5}) we see a noticeable disagreement, namely the characteristic twin peak ($\Delta_{5}$ and $\Delta_{1}$ as labeled in Fig.~\ref{fig5}) Fe minority and majority PDOS resonances are markedly distorted and in the case of the majority interface states there is a further 0.75 eV upwards shift.   This upwards shift in the interface bulk SEP majority interface states is due entirely to the positive nature of the interface error as shown by the green potential plotted in Fig.~\ref{fig4}a. 

The $\Delta_{1}$ interface state decays slowly into the MgO barrier and the $\Delta_{5}$ interface state decays rapidly into the MgO barrier (see Sec.~\ref{sec3.2.4}).  Furthermore, the 2 eV exchange splitting between the majority and minority carriers results in half-metallic like conduction between the slowly decaying $\Delta_{1}$ interface states, which dominate the TMR and spin torque characteristics of Fe/MgO/Fe junctions.\cite{ButlerTMR,heiliger08}   
By first distorting the $\Delta_{1}$ minority/majority interface states and then shifting the majority $\Delta_{1}$ interface state by 0.75 eV,  the bulk SEP approximation introduces considerable error into the half-metallic properties of the the Fe/MgO/Fe tunneling as we show in further detail in the next section.    However, the ghost SEPs clearly (as shown in red in the second column of Fig.~\ref{fig5}) are able to almost entirely compensate the for bulk SEP PDOS interface transferability errors.   This quantitative result is achieved with only first order (spherically symmetric) ghost SEPs, where angular dependent interface ghost SEPs might be necessary for more complex heterojunction interfaces.

% .........................................................................................................................................

\begin{figure*}[t]
\centering
\includegraphics[width=7in,height=2in]{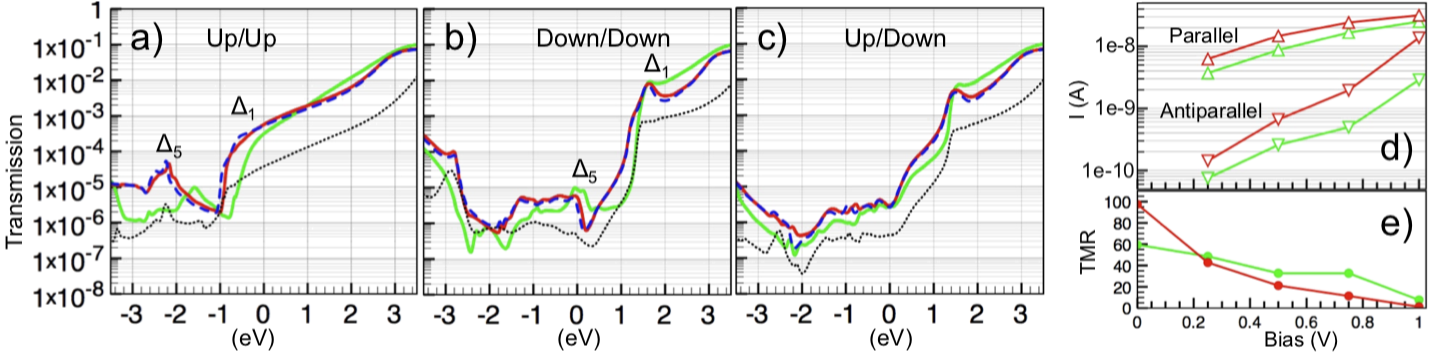}
\caption{Subfigures a), b) and c) display the parallel and anti-parallel zero bias total transmission with respect to energy through the 5 layer Fe/MgO/Fe MTJ geometry shown in Fig.~\ref{fig3}.  The self-consistent LSDA total transmission is shown in dashed dark blue, the bulk SEP result is shown in solid green, the ghost SEP result is shown in solid red, and the band gap corrected ghost SEP result is shown in dotted black.  The biased device parallel (upward pointing triangles) and anti-parallel (downward pointing triangles) currents for the ghost SEP (red solid line) and bulk SEP (green solid line) approximations are shown in subfigure d) in units of Amperes per unit cell (2.87 \AA\ by 2.87 \AA).  The voltage profile is assumed to drop linearly across the MgO barrier for the calculated $IV$ points. The ghost SEP (red solid line) and bulk SEP (green solid line) TMR ratios under bias are shown in subfigure e). The Fermi energy is set at 0 eV, the imaginary green's function broadening is set at 25 meV, and the $k$-point sampling set at 100x100 in the transverse Brillouin zone.  We define $TMR = (I_{P}-I_{AP})/I_{AP}$, where $I_{P}$ is the parallel current and $I_{AP}$ is the antiparallel current.}
\label{fig6}
\end{figure*}

\subsubsection{Fe/MgO/Fe MTJ Transmission Study}
\label{sec3.2.4}

Thus far we have performed a detailed analysis of the interface potential and PDOS errors (see Figs.~\ref{fig4} and \ref{fig5}) which result when bulk SEPs are transfered to Fe/MgO/Fe tunnel junctions.  Yet, for the purposes of electron device modeling we are most interested in the transport implications of such interface errors.   In this regard, previous studies have shown resonant interface states can significantly influence the transmission\cite{InterfaceTheoryTMR1,InterfaceTheoryTMR2, InterfaceTheoryTMR3} and spin torque\cite{SayeefSpinTorque, MathonSpinTorque, heiliger08} properties of MTJ barriers.    

The zero-bias total transmission of our Fe/MgO/Fe tunneling device geometry (see Fig.~\ref{fig3}) is presented in Fig.~\ref{fig6}.    The LSDA transmission is shown in dashed dark blue, the bulk SEP transmission in green, the ghost SEP transmission in solid red, and the MgO band gap corrected transmission is shown in dotted black (recall the dot-dashed gold correction potentials displayed in Figs.~\ref{fig2}a and \ref{fig2}b).    The zero bias parallel orientation up spin transmission is displayed in Fig.~\ref{fig6}a, the parallel orientation down spin transmission is displayed in Fig.~\ref{fig6}b, and the antiparallel transmission is displayed in Fig.~\ref{fig6}c.    

An initial inspection of Figs.~\ref{fig6}a through \ref{fig6}c, both above and below the Fermi energy where the Fe/MgO interface states are more prominent, reveals sizable transmission corrections between the bulk SEP result (in solid green) and the ghost SEP result (in solid red) -- where the latter matches the LDSA transmission (in dashed blue) quantitatively.  The Fe/MgO/Fe transmission characteristics are determined by the rapid decay of the $\Delta_5$ interface state and slow decay of the $\Delta_1$ interface state within MgO, resulting in half-metallic like tunneling between majority and minority $\Delta_1$ Fe/MgO interface states on opposite sides of the barrier (see layer 8 in Fig.~\ref{fig5}).    In Fig.~\ref{fig6} the half-metallic like conduction is immediately evident, where at -1 eV in Fig.~\ref{fig6}a and at +1 eV in Fig.~\ref{fig6}b we see a large rise in transmission corresponding to the onset of the majority and minority spin $\Delta_1$ states respectively (also shown on layer 8 in Fig.~\ref{fig5}).    

The Fermi energy (set at 0 eV in Figs.~\ref{fig5} and \ref{fig6}) low bias TMR for the LSDA, bulk SEP and ghost SEP methods is calculated to be 99.85, 59.39, 97.47 respectively.   The bulk SEP approximation underestimates the Fe/MgO/Fe low bias TMR by 30\%, and the ghost SEP interface corrections are able to compensate quite accurately.  This low bias error is due entirely to the $\Delta_1$ majority interface state broadening/shift and as shown on layer 8 in Fig.~\ref{fig5} and in Fig.~\ref{fig6}a where the bulk SEP (solid green) transmission is shifted by 0.5 eV.    The minority spin $\Delta_5$ interface state is not significantly altered by the bulk SEP approximation (see layer 8 in Fig.~\ref{fig5}) resulting in little change in the half-metallic like transmission at the Fermi energy of the $\Delta_1$ interface states in the anti-parallel orientation (see Fig.~\ref{fig6}c).   The same holds for the parallel minority transmission although it's contribution to the low bias parallel current is negligible (see Fig.~\ref{fig6}b).

Though the minority spin $\Delta_5$ interface state transmission is not significantly altered by the bulk SEP (solid green) transferability error, the minority spin $\Delta_1$ interface state zero bias transmission is however drastically underestimated between 0 eV and 1 eV (see Fig.~\ref{fig6}b).  Similarly, we can see a notable underestimation of the majority spin bulk SEP (solid green) $\Delta_1$ zero bias transmission between -1 eV and 0 eV as shown in Fig.~\ref{fig6}a.   Given the half-metallic like spin filtering property of the MgO barrier, in which tunneling between $\Delta_1$ states dominates, these errors have a significant impact on the biased tunneling current.   Without the ghost SEP interface corrections, the parallel tunneling current is underestimated by a factor of 2 and the antiparallel current by up to an order of magnitude within the bias window of 1 V as shown in Fig.~\ref{fig6}d (compare the ghost SEP solid red and bulk SEP solid green results).   Furthermore, in Fig.~\ref{fig6}e the bulk SEP (solid green) Fe/MgO/Fe TMR displays sizable deviations from the interface corrected ghost SEP (solid red) TMR, lacking the characteristic smooth decay under bias.\cite{WaldronNanotechnology}  

Similar, though less sizable, interface errors occur when we introduce bulk SEP MgO band gap corrections to our ghost SEP Fe/MgO/Fe Hamiltonian (see dotted black transmission plots in Figs.~\ref{fig6}a through \ref{fig6}c).  The bulk SEP MgO band gap corrections are plotted in Figs.~\ref{fig2}a and Figs.~\ref{fig2}b (see gold dot-dashed lines read off the left axis).    Reaching up to 5 Bohr, these band gap corrections extend into the Fe/MgO interface and suffer the same transferability problem as the uncorrected bulk SEPs.   However, from Figs.~\ref{fig6}a through \ref{fig6}c it is evident that the primary role of the band gap correction is to lower the tunneling current by an order of magnitude (as expected).  Likewise, the band gap correct TMR ratio at 43.94 compared to the LSDA TMR ratio of 99.85, can be attributed to the reintroduced interface state errors rather than a fundamental alteration in the Fe spin filtering properties of MgO.\cite{ButlerTMR}  However, it may be necessary in future studies to simultaneously address the nature of Fe/MgO interface states and MgO exchange-correlation corrections beyond the LSDA approximation.\cite{SelfInteractionCorrectedMTJ}

\section{Summary}
\label{sec4}

We have detailed a straight forward method for extracting semi-empirical pseudopotentials from real space DFT calculations.    The method has been shown to produce accurate bulk derived spherical SEPs, matching self-consistent LSDA band structure results to within 0.1 eV.   Subsequently, we examined the transferability of bulk derived MgO and Fe SEPs to Fe/MgO/Fe tunnel junctions.  It was shown that LSDA calculated Fe/MgO interface states are not adequately described by bulk SEPs.   As a result bulk SEPs can significantly underestimate or overestimate of the spin dependent transmission through thin Fe/MgO/Fe tunnel junctions.   However, the SEP tight binding method allows characterization of both the system layer by layer PDOS and real space potential.   Primarily due to the local chemical environment change experienced by interface Mg atoms, where the number of nearest neighbors is reduced from six to five, an interface fit is required to overcome bulk transferability errors.  Therefore, we put forward the notion of ghost SEPs, not localized at an atomic site but within the interface, to parameterize DFT calculated heterojunctions.     The ghost SEPs interface parameterization was shown to completely recover the LSDA interface PDOS and Fe/MgO/Fe transmission characteristics.   In general, the results emphasize the need for separate bulk and interface parameterizations when applying tight binding methods to study transport through nanoscale heterojunctions where interface states can couple effectively.  Lastly, we note that all the parameters and the device geometry discussed herein are provided for download as supplementary material.\cite{DownloadLink}

\section*{Acknowledgment}

K. H. Bevan gratefully acknowledges financial support from NSF-NIRT (Purdue), DOE (ORNL), and NSERC of Canada (McGill).  Tony Low gratefully acknowledges financial support from SRC-NRI (Purdue).  H. Guo gratefully acknowledges financial support from NSERC of Canada, FRQNT of Quebec and CIAR (McGill).  Computational support was provided by the NSF Network for Computational Nanotechnology (Purdue).


\begin{thebibliography}{99}
\bibitem[*]{kirk} Electronic address: bevankh@ornl.gov.

% --------------------------------------------------------------------------------------------
% Introduction Section Refs.

\bibitem{ButlerTMR}
W. H. Butler, X.-G. Zhang, T. C. Schulthess and J. M. MacLaren,
%``Spin-dependent tunneling conductance of Fe|MgO|Fe sandwiches''
\textit{Phys. Rev. B}, {\bf 63}, 054416 (2001).


\bibitem{MathonTMR}
J. Mathon and A. Umerski,
%``Theory of tunneling magnetoresistance of an epitaxial Fe/MgO/Fe(001) junction'',
\textit{Phys. Rev. B}, {\bf 63}, 220403 (2001).

\bibitem{JapanTMR_MgO}
S. Yuasa, T. Nagahama, A. Fukushima, Y. Suzuki and Koji Ando,
%``Giant room-temperature magnetoresistance in single-crystal Fe/MgO/Fe magnetic tunnel junctions'',
\textit{Nature Mater.}, {\bf 3}, 868 (2004).

\bibitem{AlmadenTMRMgO}
S. S. P. Parkin, C. Kaiser, A. Panchula, P. M. Rice, B. Hughes, M. Samant and S.-H. Yang,
%``Giant tunnelling magnetoresistance at room temperature with MgO (100) tunnel barriers'',
\textit{Nature Mater.} {\bf 3}, 862 (2004).

\bibitem{SwitchingSpeedDiscussionAndTMR}
J. Akerman,
%``Toward a Universal Memory,''
\textit{Science}, {\bf 308} , 508 (2005).

\bibitem{IndustryMgORev}
J.-G. Zhua and C. Park,
%``Magnetic tunnel junctions,''
\textit{Materials Today}, {\bf 9}, 36 (2006).

\bibitem{heiliger06}
C. Heiliger, P. Zahn and I. Mertig, 
%``How many Fe layers cause TMR?,''
\textit{Material Research Society Symposium Proceedings}, {\bf Q01-02}, 941, (2006).

\bibitem{GermanyMgORev}
C. Heiligera, P. Zahna and I. Mertiga,
%``Microscopic origin of magnetoresistance,''
\textit{Materials Today}, {\bf 9}, 46 (2006).

\bibitem{heiliger08}
C. Heiliger and M. D. Stiles,
%``Ab initio studies of the spin-transfer torque in magnetic tunnel junctions,''
\textit{Phys. Rev. Lett.}, {\bf 100}, 186805, (2008).


\bibitem{SpinConvergence}
R. Zeller,
%``Spin-Polarized DFT Calculations and Magnetism'',
\textit{NIC Series: Publication Series of the John von Neumann Institute for Computing}, {\bf 31}, 419 (2006).

\bibitem{SiestaOrderN}
P. Ordejon, D. A. Drabold, M. P. Grumbach and R. M. Martin,
%``Linear system-size scaling methods for electronic-structure calculations'',
\textit{Phys. Rev. B}, {\bf 51}, 1456 (1995).

\bibitem{SayeefSpinTorque}
Sayeef Salahuddin and Supriyo Datta,
%``Self-consistent simulation of quantum transport and magnetization dynamics in spin-torque based devices'',
\textit{Appl. Phys. Lett.}, {\bf 89}, 153504 (2006).


\bibitem{AliSpin}
A. A. Yanik, G. Klimeck and S. Datta,
%``Quantum transport with spin dephasing: A nonequlibrium Green's function approach'',
\textit{Phys. Rev. B}, {\bf 76}, 045213 (2007). 

\bibitem{ExpDefectPaperTMR}
P. G. Mather, J. C. Read and R. A. Buhrman,
%``Disorder, defects, and band gaps in ultrathin (001) MgO tunnel barrier layers'',
\textit{Phys. Rev. B}, {\bf 73}, 205412 (2006).


\bibitem{TheoryDefectPaperTMR}
J. P. Velev, K. D. Belashchenko, S. S. Jaswal and E. Y. Tsymbal,
%``Effect of oxygen vacancies on spin-dependent tunneling in Fe/MgO/Fe magnetic tunnel junctions'',
\textit{Appl. Phys. Lett.}, {\bf 90}, 072502 (2007).

\bibitem{WaldronNanotechnology}
D. Waldron, L. Liu and H. Guo,
%``Ab initio simulation of magnetic tunnel junctions'',
\textit{Nanotechnology}, {\bf 18}, 424026 (2007).  

\bibitem{SelfInteractionCorrectedMTJ}
S. H. Mirhosseini, K. K. Saha, A. Ernst and J. Henk,
%``Electron correlation effects in the magnetoresistance of Fe/MgO/Fe tunnel junctions: First-principles calculations'',
\textit{Phys. Rev. B}, {\bf 78}, 012404 (2008).

\bibitem{Martin}
%``Electronic Structure: Basic Theory and Practical Methods'',
\textit{Cambridge University Press}, Cambridge (2004).

\bibitem{ZungerSEP}
L.-W. Wang and A. Zunger,
%``Local-density-derived semiempirical pseudopotentials'',
\textit{Phys. Rev. B}, {\bf 51}, 17398 (1995).

\bibitem{ZungerDot}
L.-W. Wang and Alex Zunger,
%``Pseudopotential-based multiband k?p method for ;250 000-atom nanostructure systems'',
\textit{Phys. Rev. B}, {\bf 54}, 11417 (1996).

\bibitem{CerdaEHT}
J. Cerda and F. Soria,
%``Accurate and transferable extended Huckel-type tight-binding parameters'',
\textit{Phys. Rev. B}, {\bf 61}, 7965 (2000).

\bibitem{BoykinTB}
T, B. Boykin, G. Klimeck and F. Oyafuso,
%``Valence band effective-mass expressions in the sp3d5s* empirical tight-binding model applied to a Si and Ge parametrization'',
\textit{Phys. Rev. B}, {\bf 69}, 115201 (2004).

\bibitem{KlimeckGenetic}
G. Klimeck, R. C. Bowen, T. B. Boykin, C. Salazar-Lazaro, T. A. Cwik and A. Stoica,
%``Si tight-binding parameters from genetic algorithm fitting'',
\textit{Superlattices and Microstructures},  {\bf 27},  77 (2000).

\bibitem{MacLarenTMR}
J. M. MacLaren, X.-G. Zhang, W. H. Butler and Xindong Wang,
%``Layer KKR approach to Bloch-wave transmission and reflection: Application to spin-dependent tunneling'',
\textit{Phys. Rev. B}, {\bf 59}, 5470 (1999).

\bibitem{ZhangTMR}
X.-G. Zhang, W. H. Butler and Amrit Bandyopadhyay,
%``Effects of the iron-oxide layer in Fe-FeO-MgO-Fe tunneling junctions'',
\textit{Phys. Rev. B}, {\bf 68}, 092402 (2003).

% --------------------------------------------------------------------------------------------
% Method Section Refs.


\bibitem{WaldronPRL}
D. Waldron, V. Timoshevskii, Y. Hu, K. Xia and Hong Guo,
%``First principles modeling of tunnel magnetoresistance of Fe/MgO/Fe trilayers'',
\textit{Phys. Rev. Lett.}, {\bf 97}, 226802 (2006). 

\bibitem{PseudopotentialKB}
L. Kleinman and D. M. Bylander,
%``Efficacious Form for Model Pseudopotentials'',
\textit{Phys. Rev. Lett.}, {\bf 48}, 1425 (1982).

\bibitem{SIESTA}
J.M. Soler, E. Artacho, J.D. Gale, A. Garc\'ia, J. Junquera, 
P. Ordej\'on and D. S\'anchez-Portal,
%``The Siesta method for ab initio order-N materials simulation'',
\textit{J. Phys.: Cond. Matt.}, {\bf 14}, 2745 (2002).

\bibitem{ArfkenWeber}
G. B. Arfken and H. J. Weber,
%``Mathematical Methods for Physicists, 5th Ed'',
\textit{Academic Press}, Burlington, MA, USA  (2000).


% --------------------------------------------------------------------------------------------
% Bulk Section Refs.

\bibitem{AshcroftMermin}
N. W. Ashcroft and N. D. Mermin,
%``Solid State Physics'', 
\textit{Brooks Cole}, New York (1976).


\bibitem{InterfaceTheoryTMR1}
C. Heiliger, P. Zahn, B. Y. Yavorsky and I. Mertig,
%``Influence of the interface structure on the bias dependence of tunneling magnetoresistance'',
\textit{Phys. Rev. B}, {\bf 72}, 180406(R) (2005).

\bibitem{MgOLatticeConstant}
S A Canney et al, 
%``Electronic band structure of magnesium and magnesium oxide: experiment and theory", 
 \textit{J. Phys.: Condens. Matter},  {\bf 11} 7507 (1999).

\bibitem{MgOBandGap}
U. Schonberger and F. Aryasetiawan,
%``Bulk and surface electronic structures of MgO'',
\textit{Phys. Rev. B} , {\bf 52}, 8788 (1995).

% --------------------------------------------------------------------------------------------
% Two Probe Geometry Section Refs.

\bibitem{KienleSi}
D. Kienle, K.H. Bevan, G. Liang, L. Siddiqui, J.I. Cerd\'a, A.W. Ghosh, %``Extended H\"uckel theory for band structure, chemistry, and transport. II. silicon'',
\textit{J. Appl. Phys.}, {\bf 100}, 043715 (2006).

% --------------------------------------------------------------------------------------------
% Potential Profile Section Refs.

\bibitem{MathonFeO_TMR}
J. Mathon and A. Umerski,
%``Theory of tunneling magnetoresistance in a disordered Fe/MgO/Fe(001) junction'',
\textit{Phys. Rev. B}, {\bf 74}, 140404(R) (2006). 

% --------------------------------------------------------------------------------------------
% PDOS Section Refs.

\bibitem{InterfaceTheoryTMR2}
K. D. Belashchenko, J. Velev and E. Y. Tsymbal,
%``Effect of interface states on spin-dependent tunneling in Fe/MgO/Fe tunnel junctions'',
\textit{Phys. Rev. B}, {\bf 72}, 140404(R) (2005).

% --------------------------------------------------------------------------------------------
% Transmission Section Refs.

\bibitem{InterfaceTheoryTMR3}
O. Wunnicke, N. Papanikolaou, R. Zeller, P. H. Dederichs, V. Drchal and J. Kudrnovsky
%``Effects of resonant interface states on tunneling magnetoresistance'',
\textit{Phys. Rev. B}, {\bf 65}, 064425 (2002).

\bibitem{MathonSpinTorque}
O. Wessely, D. M. Edwards and J. Mathon,
%``Quantum mechanical theory of current-induced domain wall torques in ferromagnetic materials'',
\textit{Phys. Rev. B}, {\bf 77}, 174425 (2008).


\bibitem{DownloadLink}
K. H. Bevan,
``Real space first-principles semiempirical pseudopotentials for Fe/MgO/Fe," 
https://www.nanohub.org/resources/5997/ (2008).


% ------- REF EDIT Line -----------




\end{thebibliography}
\end{document}